\documentclass[%
 reprint,
 amsmath,amssymb,
 aps,
]{revtex4-2}

\usepackage{graphicx}
\usepackage{dcolumn}
\usepackage{bm}
\usepackage[utf8]{inputenc}
\usepackage{xcolor}

\begin{document}

\preprint{APS/123-QED}

\title{Controlling Purity, Indistinguishability and Quantum Yield of Incoherently Pumped Two-Level System by Spectral Filters.}

\author{Ivan V. Panyukov}
 \affiliation{Moscow Institute of Physics and Technology, 9 Institutskiy pereulok, Dolgoprudny 141700, Moscow region, Russia;}
 \affiliation{Dukhov Research Institute of Automatics (VNIIA), 22 Sushchevskaya, Moscow 127055, Russia;}
\author{Vladislav Yu. Shishkov}
 \email{vladislavmipt@gmail.com}
 \affiliation{Dukhov Research Institute of Automatics (VNIIA), 22 Sushchevskaya, Moscow 127055, Russia;}
 \affiliation{Moscow Institute of Physics and Technology, 9 Institutskiy pereulok, Dolgoprudny 141700, Moscow region, Russia;}
 \affiliation{ Center for Photonics and Quantum Materials, Skolkovo Institute of Science and Technology, Moscow, Russia; }
 \affiliation{ Laboratories for Hybrid Photonics, Skolkovo Institute of Science and Technology, Moscow, Russia; }
\author{Evgeny S. Andrianov}
 \affiliation{Dukhov Research Institute of Automatics (VNIIA), 22 Sushchevskaya, Moscow 127055, Russia;}
 \affiliation{Moscow Institute of Physics and Technology, 9 Institutskiy pereulok, Dolgoprudny 141700, Moscow region, Russia;}
 \affiliation{ Center for Photonics and Quantum Materials, Skolkovo Institute of Science and Technology, Moscow, Russia; }
 \affiliation{ Laboratories for Hybrid Photonics, Skolkovo Institute of Science and Technology, Moscow, Russia; }

\date{\today}

\begin{abstract}
Dephasing processes significantly impact the performance of deterministic single-photon sources.
Dephasing broadens the spectral line and suppresses the indistinguishability of the emitted photons, which is undesirable for many applications, primarily for quantum computing.
We consider a light emitted by a two-level system with a pulsed incoherent pump in the presence of the spectral filter. 
The spectral filter allows control of the second-order autocorrelation function, indistinguishability, and quantum yield. 
We show that narrow spectral filters can increase the indistinguishability of the emitted light while undermining the quantum yield.
The influence of the spectral filter on the second-order correlation function depends on the duration of the pump.
When the pumping pulse is long compared to the lifetime of the two-level system, the narrow spectral filters lead to a rapid increase in the second-order autocorrelation function.
In this limit, the statistics of the light from the two-level system inherit the statistics of the incoherent pump.
In the case of the short duration of the pump pulse, it is possible to preserve single-photon properties to some degree for the sub-lifetime width of the spectral filter.
Moreover, when the light emitted by the single-photon source is used to control a quantum system, e.g., cavity, the single-photon properties of the light manifest themselves differently, depending on the response time of the quantum system.
In particular, in the case of long response time, the spectral filter with sub-lifetime width can provide the near-zero second-order autocorrelation function.

\end{abstract}

\maketitle

\section{Introduction}

Single-photon sources (SPSs) have a wide range of applications, including quantum cryptography~\cite{hughes1995quantum, beveratos2002single, lounis2005single}, quantum communication~\cite{gschrey2015highly, gazzano2013bright}, quantum computing~\cite{o2007optical, cai2013experimental, michler2017quantum, o2007optical, lanyon2010towards}, quantum metrology~\cite{von2019quantum, motes2016efficient}, quantum information processing~\cite{fortsch2013versatile, babinec2010diamond}, and are compatible with integral nanophotonics~\cite{elshaari2017chip, singh2019quantum}.

Three metrics characterize the SPS performance: quantum yield, indistinguishability, and second-order autocorrelation function (purity).
The quantum yield stays for the probability of the photon emission of the SPS.
Indistinguishability indicates the ability of the emitted photons to interfere.
The measurement of the indistinguishability requires the Hong-Ou-Mandel set up~\cite{bouchard2020two, lang2013correlations, lopes2015atomic, lewis2016proposal}.
The second-order autocorrelation function determines the fluctuations of the radiation intensity and statistics of the emitted light.
Generally, the second-order autocorrelation is equal to or greater than zero.
If an SPS emits only one photon at a time, the second-order autocorrelation function is zero.
While some applications of SPS, such as quantum cryptography, requires only a low second-order autocorrelation function, most applications of SPS require appropriate values of all the above-described parameters simultaneously.

Quantum dots, SiV-, NV-centers, and single molecules serve as SPSs operating at room temperature.
However, these systems have a high level of dephasing~\cite{saxena2019improving, borri2001ultralong, kako2006gallium}, and, therefore, the indistinguishability of the emitted photons is low.
Indeed, according to~\cite{bylander2003interference}, the indistinguishability of photons emitted by SPS is equal to $I=\gamma_{\rm diss}/\left(\gamma_{\rm diss}+\gamma_{\rm deph}\right)$, where $\gamma_{\rm diss}$ and $\gamma_{\rm deph}$ are the rates of dissipation and dephasing, respectively.
For quantum dots, SiV-, NV-centers $\gamma_{\rm deph}$ can be as high as $10^5 \gamma_{\rm diss}$ at room temperature~\cite{grange2015cavity} that results in low indistinguishability.

Low indistinguishability is undesirable for many applications of SPS.
Several approaches to increase indistinguishability have emerged in recent years.
One approach is to lower the temperature of the SPS, decreasing $\gamma_{\rm deph}$~\cite{borri2005exciton, thoma2016exploring}.
Another approach is to place photon emitter near a cavity, increasing $\gamma_{\rm diss}$~\cite{grange2015cavity, choi2019cascaded, englund2005controlling, dietrich2016gaas, grange2015cavity, kiraz2004quantum, hennessy2007quantum, kaer2013microscopic, saxena2019improving}.
The achievement of high values of indistinguishability usually requires the combination of these two methods~\cite{liu2018high}.
The spectral filters can increase the indistinguishability of photons emitted by an SPS~\cite{sun2009indistinguishability}.
However, this method decreases the quantum yield.
In addition, it is unknown how the spectral filter affects the second-order autocorrelation function for SPS with a pulsed incoherent pumping.

In this paper, we consider a two-level system (TLS) with pulsed incoherent pumping as an SPS.
We investigate the effect of spectral filtering on the light emitted by an SPS.
The spectral filters affect the quantum yield, the indistinguishability, and the second-order autocorrelation function.
The sub-lifetime spectral filters provide high indistinguishability.
However, the narrow spectral filter to the light emitted by TLS leads to different second-order correlation functions depending on the duration of the pumping pulse.
When the pumping duration exceeds the lifetime of an SPS, the second-order autocorrelation function is approximately two for narrow filters.
At the short duration of the pump, a second-order autocorrelation function is low (from $1/5$ to $2/3$) for narrow spectral filters.
We also show that the statistics of the light emitted by the SPS manifests itself non-trivially when this light is applied to control a quantum-mechanical system, e.g., quantum detector.
In particular, in the case of a long response time of the quantum mechanical system controlled by the light from SPS, the second-order autocorrelation function is inversely proportional to the dephasing rate of the SPS.

\section{Model Decsription} \label{ModelDescription}
We consider a TLS as an SPS.
The Hamiltonian of this system is
\begin{equation}
	\hat{H}=\hbar\omega_0\hat \sigma^\dag\hat\sigma.
\end{equation}
where $\omega_0$ is the transition frequency, $\hat{\sigma} = |g\rangle\langle e|$ with the ground state and the excited state of the TLS $|g\rangle$ and $|e\rangle$.
Below we also refer to the state $| e \rangle$ as the working level of the SPS.
The density matrix $\hat\rho$ describes the state of the TLS and evolves according to the Lindblad equation~\cite{carmichael2013statistical, breuer2002theory}
\begin{equation} \label{Lindblad_TLS}
	\frac{\partial \hat{\rho}}{\partial t}=-\frac{i}{\hbar}\left[\hat{H},\hat{\rho}\right]+L_{\mathrm{pump}}\left[\hat{\rho}\right]
	+L_{\mathrm{diss}}\left[\hat{\rho}\right]+L_{\mathrm{deph}}\left[\hat{\rho}\right],
\end{equation}
where the relaxation operators $L_{\mathrm{pump}}\left[\hat{\rho}\right]$, $L_{\mathrm{diss}}\left[\hat{\rho}\right]$ and $L_{\mathrm{deph}}\left[\hat{\rho}\right]$ describe the incoherent pumping, energy dissipation, and dephasing with the corresponding rates $\gamma_{\mathrm{pump}}$, $\gamma_{\mathrm{diss}}$, $\gamma_{\mathrm{deph}}$
\begin{equation} \label{RelaxationOperatorsPump}
	L_{\mathrm{pump}}\left[\hat{\rho}\right]=\frac{\gamma_{\mathrm{pump}}\left(t\right)}{2}(2\hat{\sigma}^{\dagger}\hat{\rho}\hat{\sigma} -\hat{\sigma}\hat{\sigma}^{\dagger}\hat{\rho}-\hat{\rho}\hat{\sigma}\hat{\sigma}^{\dagger}),
\end{equation}
\begin{equation} \label{RelaxationOperatorsDiss}
	L_{\mathrm{diss}}\left[\hat{\rho}\right]=\frac{\gamma_{\mathrm{diss}}}{2}(2\hat{\sigma}\hat{\rho}\hat{\sigma}^{\dagger}-\hat{\sigma}^{\dagger}\hat{\sigma}\hat{\rho}-\hat{\rho}\hat{\sigma}^{\dagger}\hat{\sigma}),
\end{equation}
\begin{equation} \label{RelaxationOperatorsDeph}
	L_{\mathrm{deph}}\left[\hat{\rho}\right]=\frac{\gamma_{\mathrm{deph}}}{2}(2\hat{\sigma}^{\dagger}\hat{\sigma}\hat{\rho}\hat{\sigma}^{\dagger}\hat{\sigma}-\hat{\sigma}^{\dagger}\hat{\sigma}\hat{\rho}-\hat{\rho}\hat{\sigma}^{\dagger}\hat{\sigma}).
\end{equation}
We consider an SPS with pulse incoherent pumping.
Many experimental setups incorporate this pumping scheme~\cite{laucht2009dephasing, kiraz2004quantum, senellart2017high, hughes2013phonon, hughes2015crystal}.  
We assume that until $t=0$ TLS is in the ground state.
At the moment $t=0$, incoherent pumping begins to act on the TLS.
The incoherent pumping starts at $t=0$, then remains constant, and terminates at $t=T$, i.e. $\gamma_{\rm{pump}}\left(t\right)$ in Eq.~(\ref{RelaxationOperatorsPump}) equals $\gamma_{\rm{pump}} \neq 0$ at $0\leqslant t\leqslant T$, and is zero at other times.
In the incoherent pumping scheme, the initial pulse excites electrons to the levels of the SPS higher than the working level.
The electrons on these levels are sometimes called hot electrons.
The initial excitation then transits to the working level of the SPS~\cite{laucht2009dephasing}.
Eq.~(\ref{RelaxationOperatorsPump}) describes this process.
This model is valid for quantum dots~\cite{buckley2012engineered, strauf2007high, hanschke2018quantum, michler2000quantum, fattal2004quantum, ollivier2020reproducibility}, single molecules~\cite{steiner2007highly, lombardi2020molecule}, NV- and SiV- centers~\cite{aharonovich2011diamond, albrecht2013coupling, faraon2012coupling, ampem2009nano,babinec2010diamond, benedikter2017cavity}, two-dimensional materials~\cite{toth2019single,reserbat2021quantum}.
In general, the duration of the incoherent pumping, $T$, is the sum of two times: the pumping duration and the relaxation time from the higher levels of the SPS to the working level.
The letter time determines the lower bound of $T$ and may vary depending on the physical realization of the SPS.
For instance, in quantum dots the characteristic ratio between $\gamma_{\rm diss}$ and the relaxation rate of hot electrons towards the working level ranges from $10^2$ to $10^4$~\cite{klimov2000optical, nozik2001spectroscopy, lenngren2016hot, wang2019observation, rabouw2015dynamics, vzidek2014electron, spoor2017broadband}.
Below we refer to the time $T$ as the pumping pulse duration and mind its lower bound.

At $0\leqslant t\leqslant T$ the relaxation rate of the population of the TLS, $\gamma$ (longitudinal relaxation), and the relaxation rate of the dipole moment of the TLS, $\Gamma$ (transverse relaxation) depends both on dissipation and pumping processes and can be found from the master equation~(\ref{Lindblad_TLS})~\cite{haken1985laser,carmichael2013statistical}
\begin{equation} \label{Longitudinal_relaxation_rate}
	\gamma=\gamma_{\rm pump}+\gamma_{\rm diss},
\end{equation}
\begin{equation} \label{Transverse_relaxation_rate}
	\Gamma=\frac{\gamma_{\rm pump}+\gamma_{\rm diss}+\gamma_{\rm deph}}{2}.
\end{equation}
The physical reason for dephasing is the interaction between electrons and phonons.
In this paper, we consider the phonons in the Markovian approximation.

The metrics of SPS can be obtained from the annihilation operator of the electric field at the detector $\hat E(t)$ in the Heisenberg representation.
The dipole moment operator of the TLS ${\bf{\hat d}}$ is related to $\hat \sigma$ as ${\bf{\hat d}} = {{\bf{d}}_{eg}}\left( {\hat \sigma  + {{\hat \sigma }^\dag }} \right)$~\cite{scully1997quantum}, where ${{\bf{d}}_{eg}}$ is the matrix element of the dipole transition between the ground state and the excited state.
Thus, $\hat \sigma$ is the dimensionless dipole moment operator.
Therefore, for the light emitted by an SPS, 
\begin{equation} \label{E_sigma}
\hat{E}(t)\propto\hat{\sigma}(t-r/c), 
\end{equation}
where $r$ is the distance from the SPS to the detector~\cite{carmichael2009open}.
Hereafter we neglect the propagation time $r/c$.

A complex transmission function $F(\omega)$ completely characterizes a spectral filter.
The filter acts on the light by multiplying the Fourier component of the electric field with frequency $\omega$ by $F(\omega)$.
Thus, in the presence of a spectral filter, the parameters of an SPS are determined by the filtered electric field 
\begin{equation}
	\hat E_F(t) = \int_{-\infty}^{+\infty}\frac{d\omega}{2\pi} F(\omega)\int_{-\infty}^{+\infty}dt'\hat E(t')e^{-i\omega(t-t')}.
\end{equation}

The physical limitation on the spectral filter is the fulfillment of the causality principle for the transmitted light, which implies the Kramers--Kronig relation for $F (\omega)$.
In this case, we can introduce the transfer function
\begin{equation} \label{transfer_function}
	f(\Delta t)=\int_{-\infty}^{+\infty} F(\omega)e^{-i\omega\Delta t}\frac{d\omega}{2\pi}
\end{equation}
and express $\hat E_F (t)$ as
\begin{equation} \label{E_F}
	\hat E_F(t) = \int_{-\infty}^{t} f(t-t')\hat E(t')dt'.
\end{equation}
Below, we consider a Lorentz spectral filter
\begin{equation}
	F(\omega)=\frac{\gamma_F}{\omega-\omega_F+i\gamma_F} .
\end{equation}
The normalization $\left | F(\omega_F) \right |=1$ means that the filter transmits all the light at the frequency, $\omega_F$.
Hereafter, we assume that the filter transmits all the light at the transition frequency of the TLS, $\omega_F=\omega_0$.

\section{Indistinguishability}
The indistinguishability of the photons determines their ability to interfere.
In the case of a spectral filter, the indistinguishability is~\cite{kaer2013role}
\begin{equation} \label{I_F}
	I_{F}=\frac{\int_{-\infty}^{+\infty}dt_1\int_{-\infty}^{+\infty}dt_2\left | \left \langle \hat{E}^{\dagger}_{F}(t_1)\hat{E}_{F}(t_2) \right \rangle \right |^2}{\left ( \int_{-\infty}^{+\infty}dt\left \langle \hat{E}^{\dagger}_{F}(t)\hat{E}_{F}(t) \right \rangle\right )^2}.
\end{equation}
We present $I_{F}$ explicitly in terms of operators $\hat \sigma$ and $\hat \sigma^\dag$ in Appendix~\ref{appendix:explicit_expressions}. 
The calculation $I_{F}$ requires $\left\langle\hat{\sigma}^{\dagger}(t_1)\hat{\sigma}(t_2)\right\rangle$. 
Eq.~(\ref{Lindblad_TLS}) allows calculating this correlator via the quantum regression theorem~\cite{breuer2002theory}.

\begin{figure}
	\includegraphics[width=1\linewidth]{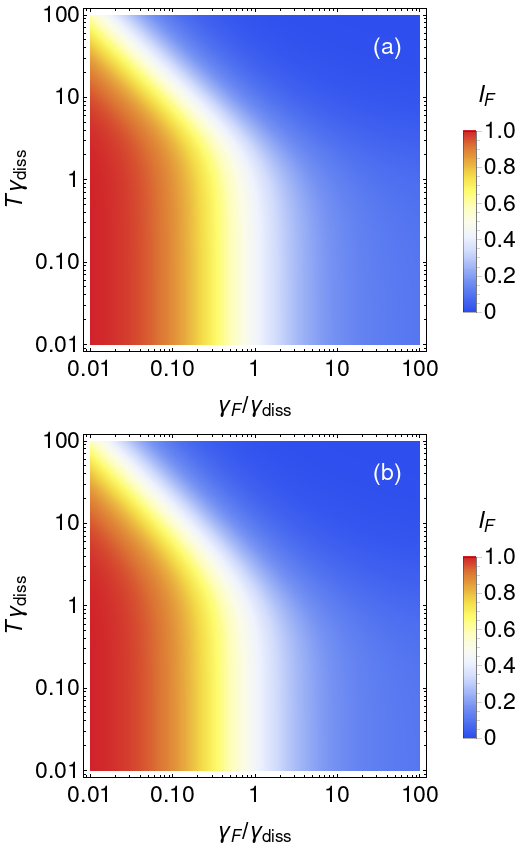}
	\caption{The dependence of the indistinguishability $I_F$ on the filter width $\gamma_F$ and the pump pulse duration $T$ in the case of $\gamma_{\rm deph}=10\gamma_{\rm diss}$ if (a) $\gamma_{\rm pump}=0.01\gamma_{\rm diss}$, (b) $\gamma_{\rm pump}=5\gamma_{\rm diss}$.} \label{ind_density}
\end{figure}

{Figure~\ref{ind_density}} shows the dependence of the indistinguishability $I_{F}$ on two parameters: the filter width and the duration of the pumping pulse.
Spectral filtering increases indistinguishability.
In the case of $T < \gamma_{\rm diss}^{-1}$, the indistinguishability increases from $\gamma_{\rm diss}/\left(\gamma_{\rm diss}+\gamma_{\rm deph}\right)$ to $1$, as the filter width decreases.
The area of high indistinguishability is wider for small pumping pulse duration.

The analytical expression for the indistinguishability in the presence of the spectral filter is lengthy.
Therefore, we consider in detail only some limiting cases.

In the case of wide spectral filter, $\gamma_{\rm F} \gg \gamma_{\rm deph}$, we obtain  
\begin{equation}\label{ind_wo_filter}
	I_F \approx I_{0}\frac{2\left ( T\gamma_{\rm diss}+e^{-T\gamma_{\rm diss}}-1 \right )}{T^2\gamma_{\rm diss}^2}.
\end{equation}
where 
\begin{equation}\label{limit_ind_1}
	I_0=\frac{\gamma_{\rm diss}}{\gamma_{\rm diss}+\gamma_{\rm deph}}.
\end{equation}
One can see that $I_F\to I_0$ as $T\to 0$, which agrees with the results obtained in~\cite{bylander2003interference}.
The indistinguishability is low at $\gamma_{\rm deph} \gg \gamma_{\rm diss}$ and wide spectral filter (see~Eq.~(\ref{ind_wo_filter})).
However, the spectral filter and the short pumping pulse can significantly increase indistinguishability.
Indeed, in the case $T \ll \max \{ \gamma_{\rm diss}^{-1}, (\gamma_{\rm diss}\gamma_{\rm deph})^{-1/2}, (\gamma_{\rm diss}\gamma_F)^{-1/2} \}$ we obtain
\begin{multline}
	I_{F} \approx I_{0}
\bigg ( 
1
+
\frac{\gamma_{\rm deph}}{\gamma_{\rm diss}+2\gamma_{F}}\cdot\frac{\gamma_{\rm deph}+3\gamma_{\rm diss}+4\gamma_{F}}{\gamma_{\rm deph}+3\gamma_{\rm diss}+2\gamma_F} 
-
\\
\frac{T^2\gamma_F\gamma_{\rm diss}}{12}\cdot\frac{\gamma_{\rm diss}+\gamma_{\rm deph}+2\gamma_{F}}{\gamma_{\rm diss}+2\gamma_F}\cdot\frac{2\gamma_{\rm deph}+3\gamma_{\rm diss}+2\gamma_F}{\gamma_{\rm deph}+3\gamma_{\rm diss}+2\gamma_F} 
\bigg).
\end{multline}
Moreover, the fulfillment of the conditions ${\gamma_F\ll \gamma_{\rm diss}}$ and ${T\ll (\gamma_{\rm diss}\gamma_F)^{-1/2}}$ makes the indistinguishability of an SPS close to unity:
\begin{equation}
I_F \approx 1 - 2\frac{\gamma_F}{\gamma_{\rm diss}} - T^2 \frac{\gamma_{\rm diss}\gamma_F}{6}
\end{equation}

\section{Second-Order Autocorrelation Function}
The second-order autocorrelation function ${g^{(2)}(t,0)=\frac{\left \langle \hat{E}^{\dagger}(t)\hat{E}^{\dagger}(t)\hat{E}(t)\hat{E}(t)\right \rangle}{\left \langle \hat{E}^{\dagger}(t)\hat{E}(t) \right \rangle^2}}$, in the presence of the spectral filter changes to 
\begin{equation} \label{g2_F}
	g^{(2)}_F(t,0)=\frac{\left \langle \hat{E}_F^{\dagger}(t)\hat{E}_F^{\dagger}(t)\hat{E}_F(t)\hat{E}_F(t)\right \rangle}{\left \langle \hat{E}_F^{\dagger}(t)\hat{E}_F(t) \right \rangle^2}.
\end{equation}
This definition implies that the detector registers the light instantaneously~\cite{glauber1965optical}.
This assumption is valid in many cases.
However, if the response time of the detector, $\tau$, is non-negligible, the second-order autocorrelation function becomes~\cite{glauber1965optical}
\begin{equation} \label{g2Fdef}
	g^{(2)}_{D,F}(t, 0)=\frac{\left \langle \hat{E}^{\dagger}_{D,F}(t)\hat{E}^{\dagger}_{D,F}(t)\hat{E}_{D,F}(t)\hat{E}_{D,F}(t) \right \rangle}{\left \langle \hat{E}^{\dagger}_{D,F}(t)\hat{E}_{D,F}(t) \right \rangle^2},
\end{equation}
where
\begin{equation}
	\hat{E}_{D,F}(t)=\int_{t}^{t+\tau}\hat{E}_F(t')dt'.
\end{equation}
Thus, unlike indistinguishability, the second-order autocorrelation function depends on $\tau$.

For detectors, $\tau$ is the duration of the interaction between light and electrons.
This time differs from the exposition time of the detector.
For instance, the exposition time of the EMCCD camera is of an order of miliseconds~\cite{edgar2012imaging}.
During this time, the photocurrent is proportional to the incoming light flux.
However, $\tau$ is the duration of the photoelectric effect process, which is much shorter than the exposition time.
In what follows, we assume the exposition time equals the detector's response time and do not analyze the influence of the exposition time on the second-order autocorrelation function.

If the light emitted by SPS controls the state of a quantum system, then $\tau$ is determined by the lifetime of this quantum system.
For instance, if the light from SPS is applied to maintain a desired state of the electromagnetic field in the cavity, then $\tau$ is the dissipation time of this cavity. 
As we show below, the single-photon properties of the light from the same SPS manifest themselves differently, depending on $\tau$.

The SPS emits light during the time $\tau_{TLS}$ which is approximately ${ \tau_{TLS}\sim T+1/ \gamma_{\rm{diss}} }$.
The first term in $\tau_{TLS}$ is the duration of the incoherent pumping pulse; the second term is the time of spontaneous relaxation after the pumping pulse ends.

Below, we consider single-photon properties of the light in two limiting cases: $\tau\gg\tau_{TLS}$ and $\tau\ll\tau_{TLS}$.

\subsection{Short detector response time}
Detection is almost instantaneous in the case $\tau\ll\tau_{TLS}$.
We denote the time at which detection occurs as $t$.
We limit ourselves to the case $t=T$, corresponding to the moment in time when the probability of light emission reaches the maximum. 
Thus, we consider $g^{(2)}_F(t,0)$ at the moment $t=T$. 

We express $g^{(2)}_F(T,0)$ in terms of operators $\hat \sigma$ and $\hat \sigma^\dag$ in Appendix~\ref{appendix:explicit_expressions}. 
To obtain $g^{(2)}_F(T,0)$, it is necessary to calculate the correlation functions $\left\langle\hat{\sigma}^{\dagger}(t_1)\hat{\sigma}^{\dagger}(t_2)\hat{\sigma}(t_3)\hat{\sigma}(t_4) \right\rangle$ and $\left\langle\hat{\sigma}^{\dagger}(t_1)\hat{\sigma}(t_2) \right\rangle$ for all possible relations between the times $\{t_1,t_2,t_3,t_4\}$.
The correlation $\left\langle\hat{\sigma}^{\dagger}(t_1)\hat{\sigma}(t_2)\right\rangle$ and the correlation $\left\langle\hat{\sigma}^{\dagger}(t_1)\hat{\sigma}^{\dagger}(t_2)\hat{\sigma}(t_3)\hat{\sigma}(t_4)\right\rangle$ for normally ordered times $\{t_1,t_2,t_3,t_4\}$ can be obtained via the standard quantum regression theorem~\cite{carmichael2013statistical}.
In all other cases, the calculation of $\left\langle\hat{\sigma}^{\dagger}(t_1)\hat{\sigma}^{\dagger}(t_2)\hat{\sigma}(t_3)\hat{\sigma}(t_4) \right\rangle$ requires generalized quantum regression theorem that was developed in~\cite{blocher2019quantum}.
The details of calculations are in Appendix~\ref{appendix:details}.

{Figure~\ref{g2t_gammaf}} shows the dependence of $g^{(2)}_F(T,0)$ on two parameters: the width of the filter, $\gamma_F$, and the duration of the pump pulse, $T$.
At $T \gg \gamma_{\rm diss}^{-1}$, a monotonous decrease of $g^{(2)}_F(T,0)$ from 2 to 0 is followed by the narrowing of the spectral filter.
This case corresponds to CW-pumping regime~\cite{panyukov2022second}, therefore, for $T \gg \gamma_{\rm diss}^{-1}$
\begin{multline}
        g_F^{(2)}(T,0) 
\approx \\ \approx
\frac{2 \gamma ^2 \left(\gamma _F+\gamma  (1-2 p)^2\right) \left(\Gamma +\gamma _F\right){}}{\left(\gamma +2 \gamma
                _F\right)\left(3 \gamma  \gamma _F+2 \gamma _F^2+\gamma ^2 (1-2 p)^2\right) \left(\Gamma +3 \gamma _F\right){}},
\end{multline}
where $p=\gamma_{\rm pump}/(\gamma_{\rm diss}+\gamma_{\rm pump})$, $\gamma$ and $\Gamma$ are defined by Eq.~(\ref{Longitudinal_relaxation_rate})--(\ref{Transverse_relaxation_rate}). 
A wide filter allows almost all the light from TLS and does not distort its statistics. 
Thus, the drop in $g^{(2)}_F(T,0)$ is expected for wide filters.

\begin{figure}[h]
	\includegraphics[width=1\linewidth]{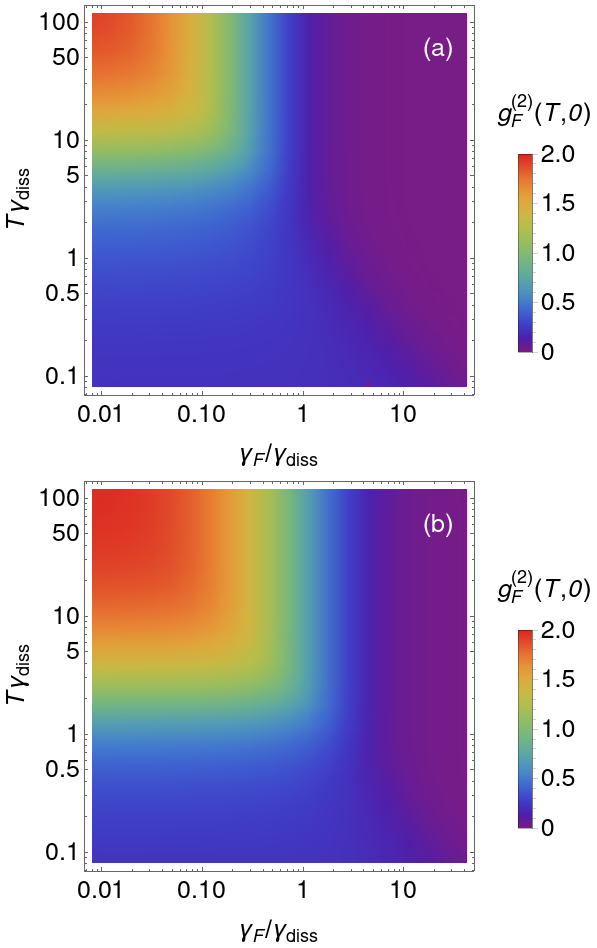}
	\caption{The dependence of $g^{(2)}_F(T,0)$ on the filter width $\gamma_F$ and the pump pulse duration $T$ at $\gamma_{\rm deph}=10\gamma_{\rm diss}$ in the case of (a) $\gamma_{\rm pump}=0.01\gamma_{\rm diss}$, (b) $\gamma_{\rm pump}=5\gamma_{\rm diss}$.} \label{g2t_gammaf}
\end{figure}

\begin{figure}[h]
	\includegraphics[width=1\linewidth]{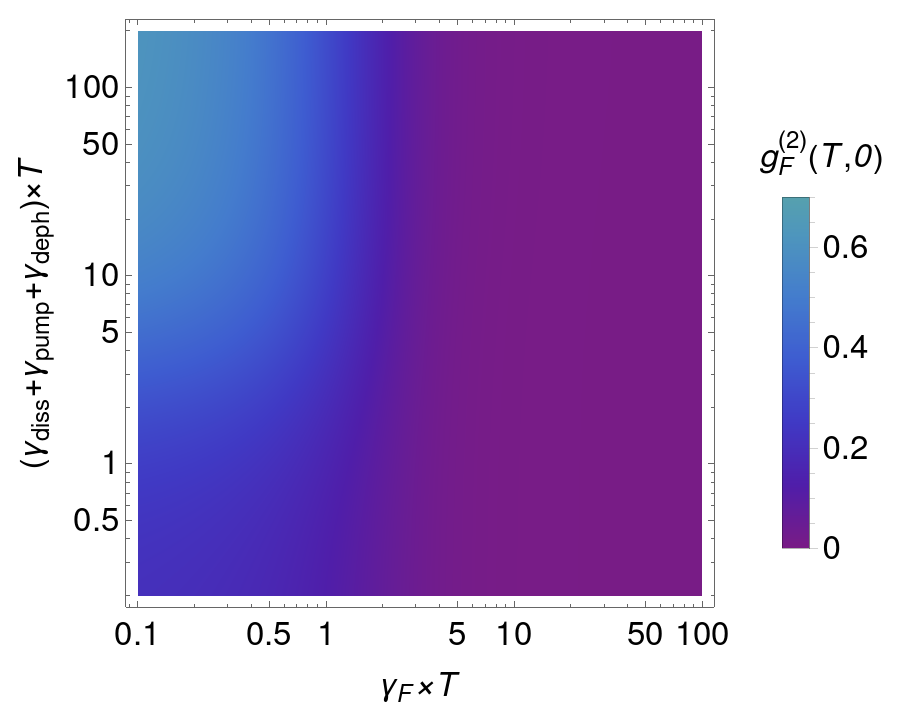}
	\caption{
The dependence of $g^{(2)}_F(T,0)$ on the filter width and the transverse relaxation rate in the limit of $T\gamma_{\rm diss}\ll 1$.
} \label{g2t_deph}
\end{figure}

{Figure~\ref{g2t_deph}} shows the dependence of $g^{(2)}_F(T,0)$, at $T\ll 1/\gamma_{\rm diss}$, on the filter width and the longitudinal relaxation rate (Eq.~(\ref{Transverse_relaxation_rate})). 
At large filter width $g^{(2)}_F(T,0)$ is zero.
However, at $T < \gamma_F^{-1}$, the second-order correlation function is in the range from $1/5$ to $2/3$ depending on the ratio between transverse relaxation time and duration of the pump pulse (Fig.~\ref{g2t_deph}). 
The naive suggestion might be $g^{(2)}_F(T,0) \approx 0$ at $T \ll \gamma_{\rm diss}$, because for fixed $\gamma_{\rm pump}$ the lower $T$ the less the probability of two consequent acts of TLS excitation and light emission.
However, {Fig.~\ref{g2t_deph}} shows that $g^{(2)}_F(T,0)$ strongly differs from zero in this limit.
We attribute this behavior to the correlations in the light emitted by TLS.
The second-order coherence function corresponding to the incoherent pump is two, which means that the dispersion of the incoherent pump is significantly above the quantum limit.
We suggest that the TLS translates this dispersion to the strong correlations of the emitted light at different moments in time, which results in the non-zero $g^{(2)}_F(T,0)$ in the limit of narrow filters and short pulse duration of the pump.
Physically, non-zero $g^{(2)}_F(T,0)$ means that during the time $\gamma_F^{-1}$ TLS has a non-zero probability of emitting two consequent photons.  
This process may happen even when $T \ll \gamma_{\rm diss}$.

\subsection{Long detector response time}
In the case $\tau\gg \tau_{TLS}$, we have ${\hat{E}_{D,F}(t)}\approx{\int_{0}^{+\infty} \hat{E}_F(t+\tau)d\tau}$. 
One can show that (see Appendix~\ref{appendix:proof})
\begin{equation} \label{equivalence_E}
\int_{0}^{+\infty} \hat{E}_F(\tau)d\tau = A\int_{0}^{+\infty} \hat{E}(\tau)d\tau,
\end{equation}
where $A=\int_{0}^{+\infty}dtf(t)=\rm{const}$.
Therefore, in the case $\tau\gg\tau_{TLS}$, the second-order autocorrelation function does not dependent on the spectral filter
\begin{multline}\label{g2_longtime}
g^{(2)}_{\infty}(0)
=\\
\frac
{
\langle
\int_{0}^{\infty}\hat E^\dag(t_1) dt_1
\int_{0}^{\infty}\hat E^\dag(t_2) dt_2
\int_{0}^{\infty}\hat E(t_3) dt_3
\int_{0}^{\infty}\hat E(t_4) dt_4
\rangle
}
{
\langle
\int_{0}^{\infty}\hat E^\dag(t_1) dt_1
\int_{0}^{\infty}\hat E(t_2) dt_2
\rangle^2
},
\end{multline}
where we assumed that the detector starts operating at $t=0$.

We express $g^{(2)}_{\infty}(0)$ in terms of operators $\hat \sigma$ and $\hat \sigma^\dag$ in Appendix~\ref{appendix:explicit_expressions}. 
To evaluate Eq.~(\ref{g2_longtime}), it is necessary to calculate the correlations $\left\langle\hat{\sigma}^{\dagger}(t_1)\hat{\sigma}(t_2) \right\rangle$ and $\left\langle\hat{\sigma}^{\dagger}(t_1)\hat{\sigma}^{\dagger}(t_2)\hat{\sigma}(t_3)\hat{\sigma}(t_4) \right\rangle$ for all possible time ratios $\{t_1,t_2,t_3,t_4\}$.
Details of calculating these correlators using the generalized quantum regression theorem are in Appendix~\ref{appendix:details}.

\begin{figure}[h]
	\includegraphics[width=1\linewidth]{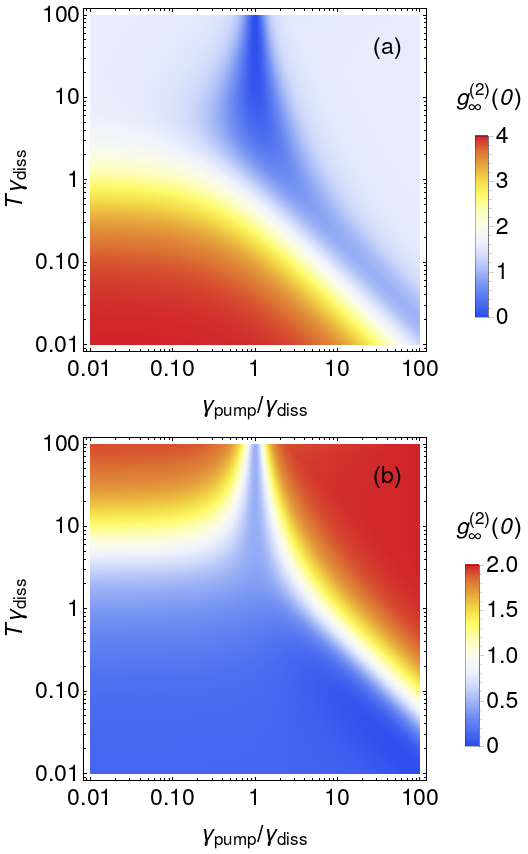}
	\caption{The dependence of $g^{(2)}_{\infty}(0)$ on the pumping speed $\gamma_{\rm pump}$ and the duration of the pump pulse $T$ at (a) $\gamma_{\rm deph}=0$, (b) $\gamma_{\rm deph}=10\gamma_{\rm diss}$.} \label{g2_inf}
\end{figure}

{Figure~\ref{g2_inf}} shows that good single-photon property is achievable at short the pumping pulse duration, $T\ll 1/\gamma_{\rm diss}$, and moderate pumping rates, $T \gamma_{\rm pump} \lesssim 1$. 
In this case, we obtain
\begin{equation} \label{g2_inf_limit}
	g^{(2)}_{\infty}(0) \approx \frac{4\gamma_{\rm diss}}{\gamma_{\rm diss}+\gamma_{\rm deph}}.
\end{equation}
Thus, at $T\ll 1/\gamma_{\rm diss}$, $T \gamma_{\rm pump} \lesssim 1$, and $\tau\gg \tau_{TLS}$ the second-order autocorrelation function decreases with increasing dephasing.
This tendency is because the light emitted by the SPS with incoherent pumping has strong correlations~\cite{panyukov2022second}.
An increase in dephasing leads to the destruction of these correlations, and as a result, $g^{(2)}_{\infty}(0)$ decreases (see~{Figure~\ref{g2_inf}}).

\section{Quantum Yield}
Due to frequency filtering, not all the light from the TLS can reach the detector, which lowers the quantum yield of the SPS. 
We denote the quantum yield in the absence of a spectral filter as ${\rm QY}$, and in the presence of the spectral filter as ${\rm QY}_F$.
The ratio between ${\rm QY}_F$ and ${\rm QY}$ can be calculated according to~\cite{kaer2013role}
\begin{equation} \label{QY}
\frac{ {\rm QY}_F }{\rm QY}
=
\frac{\int_{-\infty}^{+\infty}dt\left \langle \hat{E}^{\dagger}_{F}(t)\hat{E}_{F}(t) \right \rangle}{\int_{-\infty}^{+\infty}dt\left \langle \hat{E}^{\dagger}(t)\hat{E}(t) \right \rangle}.
\end{equation}
We present ${ {\rm QY}_F }/{\rm QY}$ in terms of operators $\hat \sigma$ and $\hat \sigma^\dag$ in Appendix~\ref{appendix:explicit_expressions}. 
To obtain ${ {\rm QY}_F }/{\rm QY}$, it is necessary to calculate $\left\langle\hat{\sigma}^{\dagger}(t_1)\hat{\sigma}(t_2) \right\rangle$.
This calculation requires the quantum regression theorem~\cite{breuer2002theory, carmichael2009open}.

\begin{figure}[h]
	\includegraphics[width=1\linewidth]{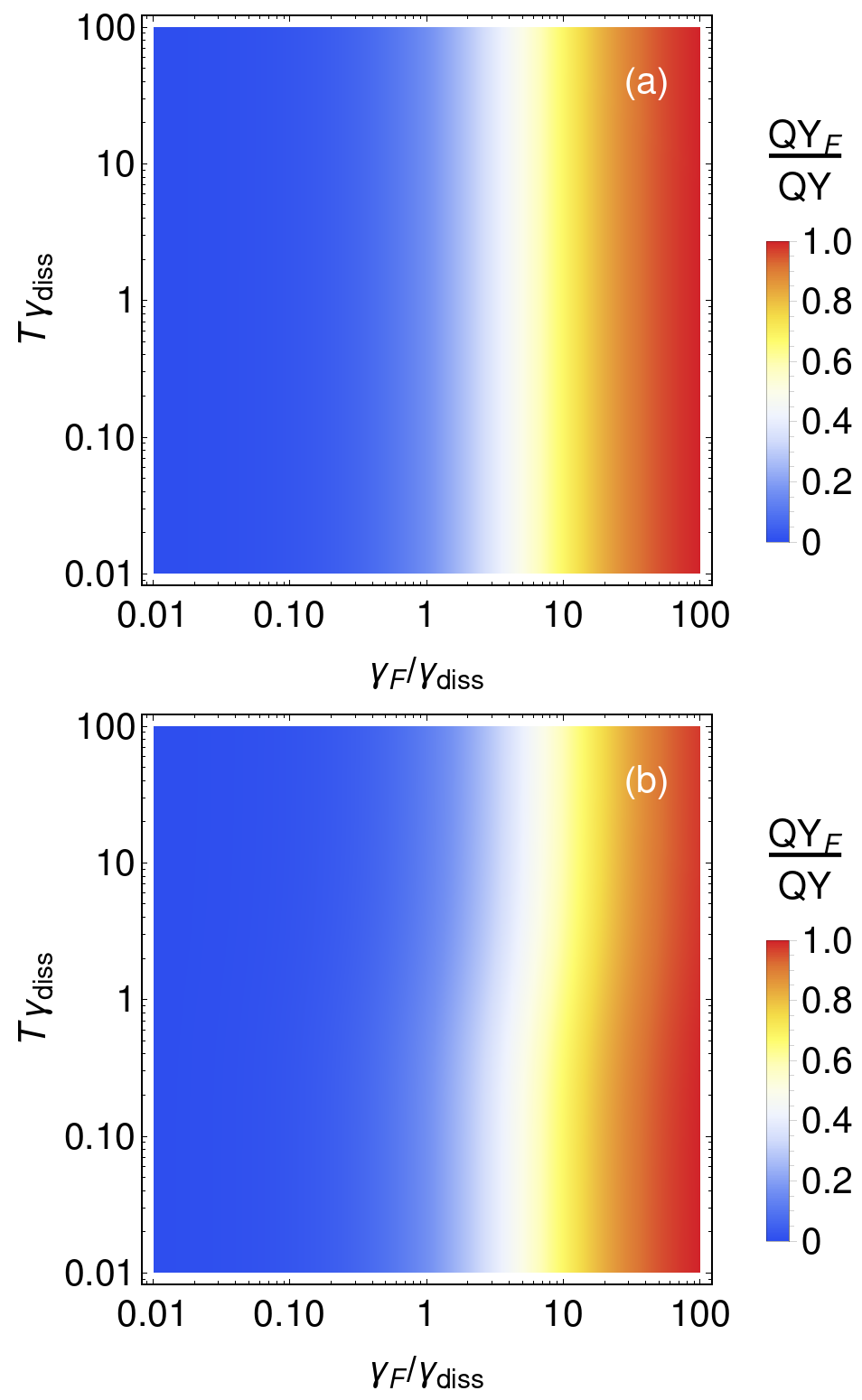}
	\caption{The dependence of the quantum yield $\eta_F$ on the filter width $\gamma_F$ and the pump pulse duration $T$ at $\gamma_{\rm deph}=10\gamma_{\rm diss}$ in the case (a) $\gamma_{\rm pump}=0.01\gamma_{\rm diss}$, (b) $\gamma_{\rm pump}=5\gamma_{\rm diss}$.} \label{eff_density}
\end{figure}

At short pumping pulse duration, $T \ll \gamma_{\rm diss}^{-1}$, the decrease in quantum yield does not depend on the $\gamma_{\rm pump}$
\begin{equation} \label{eff_T_ll}
\frac{ {\rm QY}_F }{\rm QY}
\approx
\frac{2\gamma_F}{\gamma_{\rm diss}+\gamma_{\rm deph}+2\gamma_F}.
\end{equation}
When the duration of the pumping pulse is long enough, $T \gg \gamma_{\rm diss}^{-1}$, we obtain the approximate expression for the decrease in quantum yield 
\begin{equation} \label{eff_T_gg}
\frac{ {\rm QY}_F }{\rm QY}
\approx
\frac{2\gamma_F}{\gamma_{\rm diss}+\gamma_{\rm deph}+\gamma_{\rm pump}+2\gamma_F}.
\end{equation}

{Figure~\ref{eff_density}} shows the dependence of the quantum yield on the spectral filter width and the duration of the pump pulse.
In the presence of the spectral filter, the quantum yield of the SPS drops from one to zero when the filter width decreases.
The sharp decrease in ${ {\rm QY}_F }/{\rm QY}$ occures at $\gamma_F \sim \gamma_{\rm diss}$.
Form Eq.~(\ref{Transverse_relaxation_rate}), it follows that the transverse relaxation rate of the TLS grows as the pumping rate increases. 
This trend leads to an increase in the spectral width of the light emitted by the TLS.
Therefore, when the pumping pulse is long enough, an increase in the rate of incoherent pumping with a fixed spectral filter leads to a decrease in quantum efficiency.

\section{CONCLUSION}

In this paper, we investigated the possibility of using a spectral filter to control the purity, indistinguishability, and quantum yield of the light emitted by SPS. 
We showed that it is possible to achieve near-unity indistinguishability even at fast dephasing rates by applying the spectral filter with small width $\gamma_F\ll\gamma_{\rm diss}$ and a short pumping pulse duration $T\ll 1/\gamma_{\rm diss}$.
In this case, the second-order autocorrelation function lies between $1/5$ and $2/3$ depending on the ratio between transverse relaxation time and the pump duration.
However, when the light emitted by the single-photon source is used to control a quantum system, the single-photon properties of the light manifest themselves differently, depending on the response time of the quantum system.
In particular, with the long response time of the quantum system affected by the light from SPS, the second-order autocorrelation function can be arbitrarily close to zero.
In this case, the higher the ratio between the dephasing and the dissipation rates, the lower the second-order autocorrelation function.
This tendency is due to the dephasing processes removing the strong correlations of the light emitted from an incoherently pumped SPS.
A compromise of high indistinguishability and high photon purity with the effective quantum yield is difficult to achieve because narrowing the width of the spectral filter reduces the intensity of light passing through the filter. 
Thus, the significant improvement in indistinguishability and purity of the SPS by applying a spectral filter is possible only by significantly reducing the quantum yield of this SPS.

\begin{acknowledgments}
The research was financially supported by a grant from Russian Science Foundation (project No. 20-72-10057).
V.Yu.Sh. thanks Foundation for the Advancement of Theoretical Physics and Mathematics ``Basis''.
\end{acknowledgments}

\appendix

\section{Expressions for $I_F$, $g^{(2)}_{F}(T, 0)$, $g^{(2)}_{\infty}(0)$, and ${ {\rm QY}_F }/{\rm QY}$ through $\hat \sigma$} \label{appendix:explicit_expressions}
To express $I_F$, $g^{(2)}_{F}(T, 0)$, $g^{(2)}_{\infty}(0)$, and ${ {\rm QY}_F }/{\rm QY}$ in turms of TLS operators we use use Eq.~(\ref{E_F}) and the connection between $\hat E(t)$ and $\hat \sigma (t)$ given by Eq.~(\ref{E_sigma}).
In the expressions below, we also assume that until the moment $t=0$ the TLS is in the ground state.
From~Eq.~(\ref{I_F}) we obtain 
\begin{widetext} 
\begin{equation} \label{ind}
	I_{F}=\frac{\int_{0}^{+\infty}dt_1\int_{0}^{+\infty}dt_2\int_{0}^{+\infty}dt_3\int_{0}^{+\infty}dt_4 x(t_1-t_3)x^{*}(t_2-t_4)\left \langle \hat{\sigma}^{\dagger}(t_1)\hat{\sigma}(t_2) \right \rangle\left \langle \hat{\sigma}^{\dagger}(t_3)\hat{\sigma}(t_4) \right \rangle^{*}}{\left (  \int_{0}^{+\infty}dt_1\int_{0}^{+\infty}x(t_1-t_2)\left  \langle \hat{\sigma}^{\dagger}(t_1)\hat{\sigma}(t_2) \right \rangle\right )^2}.
\end{equation}
where we introduced the notation
\begin{equation}
	x(\Delta t)=\int_{-\infty}^{+\infty}\frac{d\omega}{2\pi}\left | F(\omega) \right |^2 e^{-i\omega\Delta t}.
\end{equation}
From~Eq.~(\ref{g2_F}) and Eq.~(\ref{g2_longtime}) we obtain 
\begin{equation}\label{g2_F_sigma}
	g^{(2)}_F(t,0)=
	\frac{\int_{0}^{t}dt_1\int_{0}^{t}dt_2\int_{0}^{t}dt_3\int_{0}^{t}dt_4f^*(t-t_1)f^*(t-t_2)f(t-t_3)f(t-t_4)\langle\hat \sigma^\dag(t_1)\hat \sigma^\dag(t_2)\hat \sigma(t_3)\hat \sigma(t_4)\rangle}
	{\left(\int_{0}^{t}dt_1\int_{0}^{t}dt_2f^*(t-t_1)f(t-t_2)\langle\hat \sigma^\dag(t_1)\hat \sigma(t_2)\rangle \right)^2}.
\end{equation}
\begin{equation}\label{g2_longtime_sigma}
	g^{(2)}_{\infty}(0)=\frac{\int_{0}^{+\infty}dt_1\int_{0}^{+\infty}dt_2\int_{0}^{+\infty}dt_3\int_{0}^{+\infty}dt_4\left \langle \hat{\sigma}^{\dagger}(t_1)\hat{\sigma}^{\dagger}(t_2)\hat{\sigma}(t_3)\hat{\sigma}(t_4) \right \rangle}{\left ( \int_{0}^{+\infty}dt_1\int_{0}^{+\infty}dt_2\left \langle \hat{\sigma}^{\dagger}(t_1)\hat{\sigma}(t_2) \right \rangle \right )^2}.
\end{equation}
From~Eq.~(\ref{E_F}), we obtain
\begin{equation}\label{eff_sigma}
\frac{ {\rm QY}_F }{\rm QY} 
= \\ =
\frac{\int_{0}^{+\infty}dt\int_{0}^{t}dt_1\int_{0}^{t}dt_2f^{*}(t-t_1)f(t-t_2)\left \langle \hat{\sigma}^{\dagger}(t_1)\hat{\sigma}(t_2) \right \rangle}{\int_{0}^{+\infty}dt\left \langle \hat{\sigma}^{\dagger}(t)\hat{\sigma}(t) \right \rangle}.
\end{equation}

\section{Proof of~Eq.~(\ref{equivalence_E})} \label{appendix:proof}
To proof Eq.~(\ref{equivalence_E}) we use Eq.~(\ref{E_F}) and the initial conditions described in Section~\ref{ModelDescription}
\begin{equation} \label{proof}
\int_{0}^{+\infty} \hat{E}_F(t)dt
=\int_{0}^{+\infty}dt\int_{0}^{t}dt'f(t-t')\hat{E}(t')
=\int_{0}^{+\infty}dt'\hat{E}(t')\int_{t'}^{+\infty}dtf(t-t')
=\int_{0}^{+\infty}f(t)dt\cdot\int_{0}^{+\infty}\hat{E}(t')dt'
\end{equation}

\section{Calculation of $\left \langle \hat{\sigma}^{\dagger}(t_1)\hat{\sigma}^{\dagger}(t_2)\hat{\sigma}(t_3)\hat{\sigma}(t_4) \right \rangle$} \label{appendix:details}
Correlator $\left\langle\hat{\sigma}^{\dagger}(t_1)\hat{\sigma}^{\dagger}(t_2)\hat{\sigma}(t_3)\hat{\sigma}(t_4) \right\rangle$ with an arbitrary ratio of times can be calculated analytically using a generalized quantum regression theorem~\cite{blocher2019quantum}.
For $t_1,\,t_2,\,t_3,\,t_4\le T$, the exact expression for the correlation is $\left\langle\hat{\sigma}^{\dagger}(t_1)\hat{\sigma}^{\dagger}(t_2)\hat{\sigma}(t_3)\hat{\sigma}(t_4) \right\rangle$ is given in the Supporting Information to the article~\cite{panyukov2022second}.
For the remaining time relations $t_1,\,t_2,\,t_3,\,t_4$ and $T$ explicit expressions for $\left\langle\hat{\sigma}^{\dagger}(t_1)\hat{\sigma}^{\dagger}(t_2)\hat{\sigma}(t_3)\hat{\sigma}(t_4) \right\rangle$ are given below
\begin{equation}
\left \langle \hat{\sigma}^{\dagger}(t_1)\hat{\sigma}^{\dagger}(t_2)\hat{\sigma}(t_3)\hat{\sigma}(t_4) \right \rangle
=
\left \langle \hat{\sigma}^{\dagger}(t_1)\hat{\sigma}^{\dagger}(t_2)\hat{\sigma}(t_3)\hat{\sigma}(T) \right \rangle
e^{-(\gamma_{\rm diss}+\gamma_{\rm deph}+2i\omega_0)(t_4-T)/2}
,\;\;
t_1,\,t_2,\,t_3 \le T,
\;\;
t_4 > T,
\end{equation}
\begin{equation}
\left \langle \hat{\sigma}^{\dagger}(t_1)\hat{\sigma}^{\dagger}(t_2)\hat{\sigma}(t_3)\hat{\sigma}(t_4) \right \rangle
=
\left \langle \hat{\sigma}^{\dagger}(t_1)\hat{\sigma}^{\dagger}(t_2)\hat{\sigma}(T)\hat{\sigma}(t_4) \right \rangle
e^{-(\gamma_{\rm diss}+\gamma_{\rm deph}+2i\omega_0)(t_3-T)/2}
,\;\;
t_1,\,t_2,\,t_4 \le T,
\;\;
t_3 > T,
\end{equation}
\begin{equation}
\left \langle \hat{\sigma}^{\dagger}(t_1)\hat{\sigma}^{\dagger}(t_2)\hat{\sigma}(t_3)\hat{\sigma}(t_4) \right \rangle
=
\left \langle \hat{\sigma}^{\dagger}(t_1)\hat{\sigma}^{\dagger}(T)\hat{\sigma}(t_3)\hat{\sigma}(t_4) \right \rangle
e^{-(\gamma_{\rm diss}+\gamma_{\rm deph}-2i\omega_0)(t_2-T)/2}
,\;\;
t_1,\,t_3,\,t_4 \le T,
\;\;
t_2 > T,
\end{equation}
\begin{equation}
\left \langle \hat{\sigma}^{\dagger}(t_1)\hat{\sigma}^{\dagger}(t_2)\hat{\sigma}(t_3)\hat{\sigma}(t_4) \right \rangle
=
\left \langle \hat{\sigma}^{\dagger}(T)\hat{\sigma}^{\dagger}(t_2)\hat{\sigma}(t_3)\hat{\sigma}(t_4) \right \rangle
e^{-(\gamma_{\rm diss}+\gamma_{\rm deph}-2i\omega_0)(t_1-T)/2}
,\;\;
t_2,\,t_3,\,t_4 \le T,
\;\;
t_1 > T,
\end{equation}

\begin{multline}
\left \langle \hat{\sigma}^{\dagger}(t_1)\hat{\sigma}^{\dagger}(t_2)\hat{\sigma}(t_3)\hat{\sigma}(t_4) \right \rangle
=
\\
=
\left \langle \hat{\sigma}^{\dagger}(t_1)\hat{\sigma}^{\dagger}(T)\hat{\sigma}(t_3)\hat{\sigma}(T) \right \rangle
e^{-(\gamma_{\rm diss}+\gamma_{\rm deph})\left |t_2-t_4\right |/2}e^{-i\omega_0(t_4-t_2)}e^{-\gamma_{\rm diss}(\min\{t_2,t_4\}-T)}
,\;\;
t_1,\,t_3 \le T,
\;\; 
t_2,\,t_4 > T,
\end{multline}
\begin{multline}
\left \langle \hat{\sigma}^{\dagger}(t_1)\hat{\sigma}^{\dagger}(t_2)\hat{\sigma}(t_3)\hat{\sigma}(t_4) \right \rangle
=
\\
=
\left \langle \hat{\sigma}^{\dagger}(T)\hat{\sigma}^{\dagger}(t_2)\hat{\sigma}(t_3)\hat{\sigma}(T) \right \rangle
e^{-(\gamma_{\rm diss}+\gamma_{\rm deph})\left |t_1-t_4\right |/2}e^{-i\omega_0(t_4-t_1)}e^{-\gamma_{\rm diss}(\min\{t_1,t_4\}-T)}
,\;\;
t_2,\,t_3 \le T,
\;\; 
t_1,\,t_4 > T,
\end{multline}
\begin{multline}
\left \langle \hat{\sigma}^{\dagger}(t_1)\hat{\sigma}^{\dagger}(t_2)\hat{\sigma}(t_3)\hat{\sigma}(t_4) \right \rangle
=
\\
=
\left \langle \hat{\sigma}^{\dagger}(T)\hat{\sigma}^{\dagger}(t_2)\hat{\sigma}(T)\hat{\sigma}(t_4) \right \rangle
e^{-(\gamma_{\rm diss}+\gamma_{\rm deph})\left |t_1-t_3\right |/2}e^{-i\omega_0(t_3-t_1)}e^{-\gamma_{\rm diss}(\min\{t_1,t_3\}-T)}
,\;\;
t_2,\,t_4 \le T,
\;\; 
t_1,\,t_3 > T,
\end{multline}
\begin{multline}
\left \langle \hat{\sigma}^{\dagger}(t_1)\hat{\sigma}^{\dagger}(t_2)\hat{\sigma}(t_3)\hat{\sigma}(t_4) \right \rangle
=
\\
=
\left \langle \hat{\sigma}^{\dagger}(t_1)\hat{\sigma}^{\dagger}(T)\hat{\sigma}(T)\hat{\sigma}(t_4) \right \rangle
e^{-(\gamma_{\rm diss}+\gamma_{\rm deph})\left |t_2-t_3\right |/2}e^{-i\omega_0(t_3-t_2)}e^{-\gamma_{\rm diss}(\min\{t_2,t_3\}-T)}
,\;\;
t_1,\,t_4 \le T,
\;\; 
t_2,\,t_3 > T,
\end{multline}
\begin{equation}
\left \langle \hat{\sigma}^{\dagger}(t_1)\hat{\sigma}^{\dagger}(t_2)\hat{\sigma}(t_3)\hat{\sigma}(t_4) \right \rangle
=
0
,\;\;
t_1,\,t_2 \le T,
\;\; 
t_3,\,t_4 > T,
\end{equation}
\begin{equation}
\left \langle \hat{\sigma}^{\dagger}(t_1)\hat{\sigma}^{\dagger}(t_2)\hat{\sigma}(t_3)\hat{\sigma}(t_4) \right \rangle
=
0
,\;\;
t_3,\,t_4 \le T,
\;\; 
t_1,\,t_2 > T,
\end{equation}

\begin{equation}
\left \langle \hat{\sigma}^{\dagger}(t_1)\hat{\sigma}^{\dagger}(t_2)\hat{\sigma}(t_3)\hat{\sigma}(t_4) \right \rangle
=
0
,\;\;
t_4 \le T,
\;\;
t_1,\,t_2,\,t_3 > T,
\end{equation}
\begin{equation}
\left \langle \hat{\sigma}^{\dagger}(t_1)\hat{\sigma}^{\dagger}(t_2)\hat{\sigma}(t_3)\hat{\sigma}(t_4) \right \rangle
=
0
,\;\;
t_3 \le T,
\;\;
t_1,\,t_2,\,t_4 > T,
\end{equation}
\begin{equation}
\left \langle \hat{\sigma}^{\dagger}(t_1)\hat{\sigma}^{\dagger}(t_2)\hat{\sigma}(t_3)\hat{\sigma}(t_4) \right \rangle
=
0
,\;\;
t_2 \le T,
\;\;
t_1,\,t_3,\,t_4 > T,
\end{equation}
\begin{equation}
\left \langle \hat{\sigma}^{\dagger}(t_1)\hat{\sigma}^{\dagger}(t_2)\hat{\sigma}(t_3)\hat{\sigma}(t_4) \right \rangle
=
0
,\;\;
t_1 \le T,
\;\;
t_2,\,t_3,\,t_4 > T,
\end{equation}
\begin{equation}
\left \langle \hat{\sigma}^{\dagger}(t_1)\hat{\sigma}^{\dagger}(t_2)\hat{\sigma}(t_3)\hat{\sigma}(t_4) \right \rangle
=
0
,\;\;
t_1,\,t_2,\,t_3,\,t_4 > T.
\end{equation}
\end{widetext}

\bibliography{TradeOff}

\end{document}